\documentclass[11pt,oneside,a4paper]{article}
\usepackage[left=1in,right=1in,top=1in,bottom=1in]{geometry}
\usepackage{amsmath,amssymb}
\usepackage[dvips]{graphicx}

\usepackage{hyperref}

\usepackage{setspace}

\usepackage{authblk}

\hypersetup{colorlinks=true, linkcolor=blue, citecolor=black, bookmarks=false, pdfstartview={FitH}}

\begin{document}

\title{\bf Low energy Greybody factors for fermions \\
emitted by a Schwarzschild-de Sitter black hole}

\author[1]{\bf Ciprian A. Sporea\thanks{ciprian.sporea89@e-uvt.ro}}
\author[2]{\bf Andrzej Borowiec\thanks{andrzej.borowiec@ift.uni.wroc.pl}}
\affil[1]{Faculty of Physics, West University of Timi\c soara, V.  P\^ arvan Ave.  4, RO-300223 Timi\c soara, Romania}
\affil[2]{Institute for Theoretical Physics, pl. M. Borna 9, 50-204, Wroclaw, Poland}

\date{\small \today}

\maketitle

\begin{abstract}
In this paper we are discussing the problem of low-energy greybody factors for fermions emitted by a Schwarzschild-de Sitter black hole. In our study we are using the analytical methods proposed by Unruh some time ago for determining the greybody factors. We have found that at low energies the greybody factors are constant for a given total angular momentum (similar to what happens in the case of scalar particles reported before in the literature). Also our results are indicating an enhancement in the energy spectrum if one is increasing the value of the cosmological constant. These results are consistent with numerical calculations performed in Ref.[14] by S.F. Wu et al. (Phys.Rev.D 78, 084010).
\ \ \\

{\bf Keywords:} Dirac fermions $\bullet$ greybody factors $\bullet$ Schwarzschild-de Sitter black hole.

{\bf PACS (2008):} 04.62.+v $\bullet$ 04.50.Gh \\

\end{abstract}

\section{Introduction}

Since the discovery more than a decade ago that our Universe has entered a new phase of accelerated expansion \cite{20,20a} the study of non-flat asymptotically space-times geometries has drawn more and more attention from the scientific community. One of the most studied non-flat geometries is the de Sitter one due to it's rich symmetries and the fact that if our Universe continues it's accelerated expansion (which could be driven by the presence in the Einstein equations of a nonzero cosmological constant) then it will pass in the far future through a de Sitter phase. Moreover, the inflationary phase \cite{21,21a} can be also described with a good approximation by a de Sitter geometry. It is worth noticing that Lagrangian modified gravity, which provides a dynamical mechanism for an effective cosmological constant implemented by the higher curvature terms in the Lagrangian, has found an application in cosmology (see e.g. \cite{CF,ABF}) in order to explain dark energy effects. Moreover, these so-called $f(R)$ theories contain Schwarzschild-de Sitter metric as a vacuum solution \cite{FFV}. They were also tested in the solar-system and galactic scales \cite{AFRT}.

On the other hand the study of black holes also continues to be a hot topic in research today. Since Hawking's \cite{22} proposal that black holes can radiate energy through emission of particles, there have been a lot of research done in this line. However, it seems that the vast majority of studies concentrated on the asymptotically flat case while the non-flat case has been less studied so far. One of the quantities characteristic to the Hawking radiation are the greybody factors that describe  the departure of the spectrum from the one of a pure black body.

Thus in the present study we investigate the problem of low energy greybody factors for fermions emitted by a Schwarzschild-de Sitter (SdS) black hole. In order to do this we apply similar methods used by Unruh \cite{1} for the case of Schwarzschild black holes (see also more recent studies \cite{Kim,RS}). This consists mainly in obtaining approximative analytical solutions to the field equations in the vicinity of the black hole, respectively the cosmological horizons and then matching these solutions (technique valid only at low energies) and determining the absorbtion probabilities or equivalently the greybody factors. Given the complexity of the differential equations in the region between the two horizons we will perform our analysis only for small values of the cosmological constant.

Up to our knowledge this is one of the first attempts to study analytically the emission of fermions from SdS black holes. Besides the numerical study performed in ref. \cite{3} we haven't found other studies that treat the problem of fermion emission in the 4-dimensional SdS geometry. The existing literature seems to be concentrated on the scalar case for which both numerical and analytical studies where performed \cite{2,3,14,16,15,15a,15b,23}; or on two-dimensional Vaidya-de Sitter toy models \cite{17,17a,17b}; on thermodynamical radiation via tunneling \cite{18} or the associated quasinormal frequencies \cite{19,19a,19b}.

The outline of the paper consists in presenting some general notions about the Dirac equation in curved space-times and a few basic proprieties of SdS black holes in section 2. The solutions to the field equations are presented in section 3 where we concentrate our attention especially on finding the Dirac modes in an intermediate region between the two horizons present in SdS geometry. Section 4 is concentrated on finding the greybody factors for this geometry followed by a brief study of energy spectrum. We end our investigation with a brief section dedicated to our main conclusions.

Trough this paper we use natural units with $c=G=\hbar=1$ and a $(+,-,-,-)$ metric signature.

\section{Preliminaries}

In this section we present a short review of the Dirac equation written in the Cartesian gauge proposed some time ago in \cite{6,6a}. The main advantage of this gauge is that it allows us to separate the spherical degrees of freedom in the same manner done for the central problems in a flat space-time \cite{7}. This implies that the Dirac field will be a linear combination of particular modes. Those of positive frequency and given energy $\epsilon$ will read
\begin{equation}\label{p1}
\Psi_{\epsilon,k,m_{j}}(t,r,\theta,\phi)=a(r)[F^{+}_{\epsilon,k}(r)\Phi^{+}_{m_{j},k}(\theta,\phi)
+F^{-}_{\epsilon,k}(r)\Phi^{-}_{m_{j},k}(\theta,\phi)]e^{-i\epsilon t}\,,
\end{equation}
where we denoted by $\Phi^{\pm}_{m_{j}, k}$ the usual four-component angular spinors \cite{7,8} and by $F^{\pm}_{\epsilon,k}(r)$ the radial part of the wave function; $a(r)$ is a metric dependent function. The parameters $m_j$ and $k$ in (\ref{p1}) stand for the standard quantum numbers and the eigenvalue $k$ is related with the total angular momentum $j$ and the orbital angular momentum $l$ by the following forlmulas
\begin{equation}\label{p2}
 j=|k|-\frac{1}{2} \qquad l=|k+\frac{1}{2}|-\frac{1}{2}
\end{equation}
In particular, $k=-1$ corresponds to $j=\frac{1}{2},\ l=0$ while $k=1$ gives rise $j=\frac{1}{2},\ l=1$.

The Dirac equation describing a free spinor field $\psi$ of mass $m$ in a curved space-time background can be defined in a local frame (characterised by the orthogonal tetrad-gauge fields $e_{\hat\alpha}$ and $\hat e^{\hat\alpha}$) as
\begin{equation}\label{p3}
 \left( i\gamma^{\hat\alpha}D_{\hat\alpha} -m \right)\psi=0\,,
\end{equation}
The covariant derivative can be defined with the help of the spin connections $S^{\hat\alpha \hat\beta}=\frac{i}{4}[\gamma^{\hat\alpha}, \gamma^{\hat\beta} ]$, where the $\gamma$-matrices satisfy  $\{ \gamma^{\hat\alpha}, \gamma^{\hat\beta} \}=2\eta^{\hat\alpha \hat\beta}$,
\begin{equation}\label{p4}
D_{\hat\alpha}=e_{\hat\alpha}^{\mu}D_{\mu}=\hat\partial_{\hat\alpha}+\frac{i}{2}S^{\hat\beta \cdot}_{\cdot
\hat\gamma}\hat\Gamma^{\hat\gamma}_{\hat\alpha \hat\beta}\,
\end{equation}
and where the local frame components of the connection read $\hat\Gamma^{\hat\sigma}_{\hat\mu \hat\nu}=e_{\hat\mu}^{\alpha} e_{\hat\nu}^{\beta} (\hat e_{\gamma}^{\hat\sigma}\Gamma^{\gamma}_{\alpha \beta}-\hat e^{\hat\sigma}_{\beta, \alpha})$, with $\Gamma^{\gamma}_{\alpha \beta}$ being the standard Christoffel symbols.

The line element $ds^2$ for a given manifold can be written in terms of the tetrad fields and the 1-forms $\omega^{\hat\mu}=\hat e^{\hat\mu}_{\nu}dx^{\nu}$ as
\begin{equation}\label{p5}
ds^2=\eta_{\hat\alpha\hat\beta}\omega^{\hat\alpha}\omega^{\hat\beta}=g_{\mu\nu}dx^{\mu}dx^{\nu}
\end{equation}

Using the tetrads defined in ref.\cite{6,6a}, namely:
\begin{equation}\label{p5a}
\begin{split}
&\omega^0=w(r)dt \,,\\
&\omega^1=\frac{w(r)}{u(r)}\sin\theta\cos\phi \,dr+ \frac{r w(r)}{v(r)}\cos\theta\cos\phi \,d\theta-\frac{r w(r)}{v(r)}\sin\theta\sin\phi \,d\phi\,, \\
&\omega^2=\frac{w(r)}{u(r)}\sin\theta\sin\phi \,dr+ \frac{r w(r)}{v(r)}\cos\theta\sin\phi \,d\theta+\frac{r w(r)}{v(r)}\sin\theta\cos\phi \,d\phi\,, \\
&\omega^3=\frac{w(r)}{u(r)}\cos\theta \,dr- \frac{r w(r)}{v(r)}\sin\theta \,d\theta\,,
\end{split}
\end{equation}
the line element in a central static chart with spherical coordinates $(t,r,\theta,\phi)$ can be expressed as
\begin{equation}\label{p6}
ds^{2}=w(r)^{2}\left[dt^{2}-\frac{dr^{2}}{u(r)^2}-
\frac{r^2}{v(r)^2}(d\theta^{2}+\sin^{2}\theta d\phi^{2})\right]\,
\end{equation}
where $u(r)$, $v(r)$ and $w(r)$ are three arbitrary $r$-dependent functions. The form of $a(r)$ introduced in (\ref{p1}) turns out to be $a(r)=v(r) w(r)^{-3/2}/r$.

For a Schwarzschild-de Sitter space-time the line element is defined by
\begin{equation}\label{p7}
ds^{2}=h(r)\,dt^{2}-\frac{dr^{2}}{h(r)}- r^{2} (d\theta^{2}+\sin^{2}\theta~d\phi^{2})\,,
\end{equation}
with $h(r)$ given by
\begin{equation}\label{p8}
h(r)=1-\frac{2M}{r}-\frac{\Lambda}{3}r^2\,,
\end{equation}
where $M$ is the mass of the black hole and $\Lambda$ stands for a positive cosmological constant.

Comparing (\ref{p6}) with (\ref{p7}) we can make the following identifications
\begin{equation}\label{p9}
u(r)=h(r), \ \ \ \ \ \ \ \ \ v(r)=\sqrt{h(r)}=w(r)
\end{equation}

It can be shown \cite{6,6a} that the radial wave functions satisfy a radial eigenvalue problem $H_r{\cal F}=\epsilon{\cal F}$ written in a matrix form as
\begin{equation}\label{sis}
\left(\begin{array}{cc}
    m\,\sqrt{h(r)}& -h(r)\frac{d}{dr}+\frac{k}{r}\sqrt{h(r)}\\
&\\
  h(r)\frac{d}{dr}+\frac{k}{r}\sqrt{h(r)}& -m\,\sqrt{h(r)}
\end{array}\right)
\left(\begin{array}{cc}
    F^{+}_{\epsilon,k}(r)\\
&\\
  F^{-}_{\epsilon,k}(r)
\end{array}\right)=\epsilon \left(\begin{array}{cc}
    F^{+}_{\epsilon,k}(r)\\
&\\
  F^{-}_{\epsilon,k}(r)
\end{array}\right)
\end{equation}

This system of first order differential equations is not analytically solvable due to the complicated form of $h(r)$. However, in specific regions of the space-time we can find approximative analytical solutions.

Before moving on to find solutions to eq. (\ref{sis}) let us say a few words about the properties of (\ref{p7}). Looking for the roots of $h(r)$ which will point out the singularities of the metric (\ref{p7}) we find that it can have in general three different roots which will correspond to three horizons. However, only two of them are physical and they correspond to the black hole horizon (located at $r=r_b$) and respectively to the cosmological horizon (located at $r=r_c$). The third solution to $h(r)=0$ turns out to be related to the other two by $r_X=-(r_b+r_c)$. Given that the condition $0<\xi\equiv\Lambda M^2<1/9$ is satisfied\cite{9,10} we will always have the two horizons satisfying  $r_b<r_c$. For the critical value of $\xi=1/9$ the two horizons will merge and a Nariai\cite{11} black hole will appear, while for $\xi>1/9$ no horizons exist. In this paper we will study only the case of $\xi<1/9$.

\section{Solutions to field equation}

The above coupled differential equations (\ref{sis}) can be transformed into a single second order equation. After making a change of variables defined by
\begin{equation}\label{ss1}
\frac{dr}{dx}=\frac{h}{1+\lambda\sqrt{h}}
\end{equation}
where $\lambda=m/\epsilon$, the new equation for the upper component $F^{+}(r)$ reads
\begin{equation}\label{ec1}
\frac{d^2F^+}{dx^2}+\left[\epsilon^2\left( \frac{1-\lambda\sqrt{h}}{1+\lambda\sqrt{h}} \right)+ \frac{d}{dx}\left( \frac{k\sqrt{h}}{(1+\lambda\sqrt{h})r} \right)- \frac{k^2h}{(1+\lambda\sqrt{h})^2\,r^2} \right]F^+=0
\end{equation}

In what follows we will solve this equation using the methods proposed by Unruh in \cite{1} and developed further, for ex. in refs. \cite{Kim,RS}. The solutions obviously depend on the value of the parameters: $(k,M,\Lambda)$. However, in order to simplify the notation we shall skip writing them explicitly.

\subsection{Solutions near the two horizons $r_b$ and $r_c$}

In the regions near the black hole horizon (located at $r\rightarrow r_b$), respectively the cosmological horizon (at $r\rightarrow r_c$) the function $h(r)$ goes to zero so that Eq. (\ref{ec1}) reduces to a more simple equation
\begin{equation}\label{s1}
\frac{d^2F^+}{dx^2}+\epsilon^2F^+=0
\end{equation}
which has the following general solution
\begin{equation}\label{s2}
F^+(x)=A\,e^{-i\epsilon x}+B\,e^{i\epsilon x}
\end{equation}

It is easy to see from (11) that the variable $x$ behaves near the two horizons as $x \sim \ln h$. Integration constant can be chosen by imposing suitable initial conditions
\begin{equation}\label{s3}
x \approx \begin{cases} \left( \frac{2M}{r^2_{b}}-\frac{2\Lambda}{3}r_{b}\right)^{-1}\ln h\equiv p\ln h, & \mbox{if } r\rightarrow r_{b} \\ \left( \frac{2M}{r^2_c}-\frac{2\Lambda}{3}r_c\right)^{-1}\ln h\equiv q\ln h, & \mbox{if } r\rightarrow r_c \end{cases}
\end{equation}

Taking into account the ingoing boundary condition according to which only ingoing modes must exist in the vicinity of the black hole horizon, the solution in this (transition) region becomes
\begin{equation}\label{s4}
F^+_b=A^{tr} e^{-i\epsilon x}\approx A^{tr} e^{-i\epsilon p\ln{h}}
\end{equation}

At the cosmological horizon we have no restriction, thus the solution will be a combination of ingoing and outgoing modes
\begin{equation}\label{s5}
\begin{split}
F^+_c&=A^{in} e^{-i\epsilon x}+A^{out} e^{i\epsilon x}\\
&\approx A^{in} e^{-i\epsilon q\ln{h}}+A^{out} e^{i\epsilon q\ln{h}}
\end{split}
\end{equation}

\subsection{Solutions in the intermediate region $r_b < r <r_c$}

If we take into consideration the fact that in the intermediate region the terms proportional to $\epsilon^2$ or $m^2$ are much smaller than the other terms\cite{1}, then Eq. (\ref{ec1}) for the radial wave function reduces to
\begin{equation}\label{s6}
\frac{d^2F^+}{dx^2}+\left[\frac{d}{dx}\left( \frac{k\sqrt{h}}{(1+\lambda\sqrt{h})r} \right)- \frac{k^2h}{(1+\lambda\sqrt{h})^2\,r^2} \right]F^+=0
\end{equation}

Our next goal will be to find an analytical solution to this equation. We start by introducing a new function $H$ defined as
\begin{equation}\label{s7}
H\equiv\frac{dF^+}{dx}+\frac{k\sqrt{h}}{(1+\lambda\sqrt{h})r}F^+
\end{equation}
with the help of which Eq. (\ref{s6}) can be rewritten as
\begin{equation}\label{s8}
\frac{dH}{dr}-\frac{k}{\sqrt{h}r}H=0
\end{equation}

Due to the complicated form of $h(r)$ Eq.(\ref{s8}) has no integrable solution. However, in the more simple case of Schwarzschild black hole for which
\begin{equation}\label{s8a}
h\rightarrow h_0=1-\frac{2M}{r}
\end{equation}
equation (\ref{s8}) admits the following solution \cite{1}
\begin{equation}\label{s9}
H_0=C\left( \frac{1-\sqrt{h_0}}{1+\sqrt{h_0}} \right)^{-k}
\end{equation}

Because the cosmological constant $\Lambda$ has for the present expansion a very small value we can consider $b\equiv\Lambda/3$ as a small parameter and expand $1/\sqrt{h}$ around the value b=0 to obtain
\begin{equation}\label{s10}
\frac{k}{\sqrt{h}r}\approx \frac{k}{\sqrt{h_0}r}+\frac{1}{2}\frac{k r}{(\sqrt{h_0})^3}\,b
\end{equation}

Considering $H$ as a function of $b$ one can expand it for small values of the parameter $b$ (till linear order) as follows
\begin{equation}\label{s11}
\begin{split}
H(r,b)&\approx H_0(r)+b\,\frac{\partial H}{\partial b}\bigg|_{b=0} \\
&=H_0(r)+b\,\delta H
\end{split}
\end{equation}

Substituting (\ref{s10}) and (\ref{s11}) into Eq. (\ref{s8}) and keeping only the first order terms in $b$ we arrive at the following equation for $\delta H$
\begin{equation}\label{s12}
\frac{d\delta H}{dr}-\frac{k}{\sqrt{h_0}r}\delta H=\frac{1}{2}\frac{k r}{(\sqrt{h_0})^3}H_0
\end{equation}
which can be immediately integrated to obtain
\begin{equation}\label{s13}
\delta H=H_0\left(\frac{1}{2}\int\frac{k r}{(\sqrt{h_0})^3}\,dr \right)
\end{equation}

Now we can write the final expresion for $H$ as
\begin{equation}\label{s14}
H=H_0\left( 1+b\,I \right)
\end{equation}
where we denoted by $I$ the following quantity
\begin{equation}\label{s15}
\begin{split}
I=\frac{1}{2}\int\frac{k r}{(\sqrt{h_0})^3}\,dr &=\frac{1}{2}\frac{k}{r-2M}\bigg[ r\sqrt{h_0}(r^2+5M\,r-30M^2) \\
& +15M^2(r-2M)\ln(2r\sqrt{h_0}+2r-2M) \bigg]
\end{split}
\end{equation}

The accuracy of such solutions can be checked by comparison with a numerical one, provided that one imposes the same initial condition: $H_{numerical}(\frac{1}{2}(r_b+r_c))=H(\frac{1}{2}(r_b+r_c))$.
The corresponding plots are collected in Fig. 1 for two different values of $\Lambda r^2_b$.

\begin{figure}[h!t]
\includegraphics[scale=0.3]{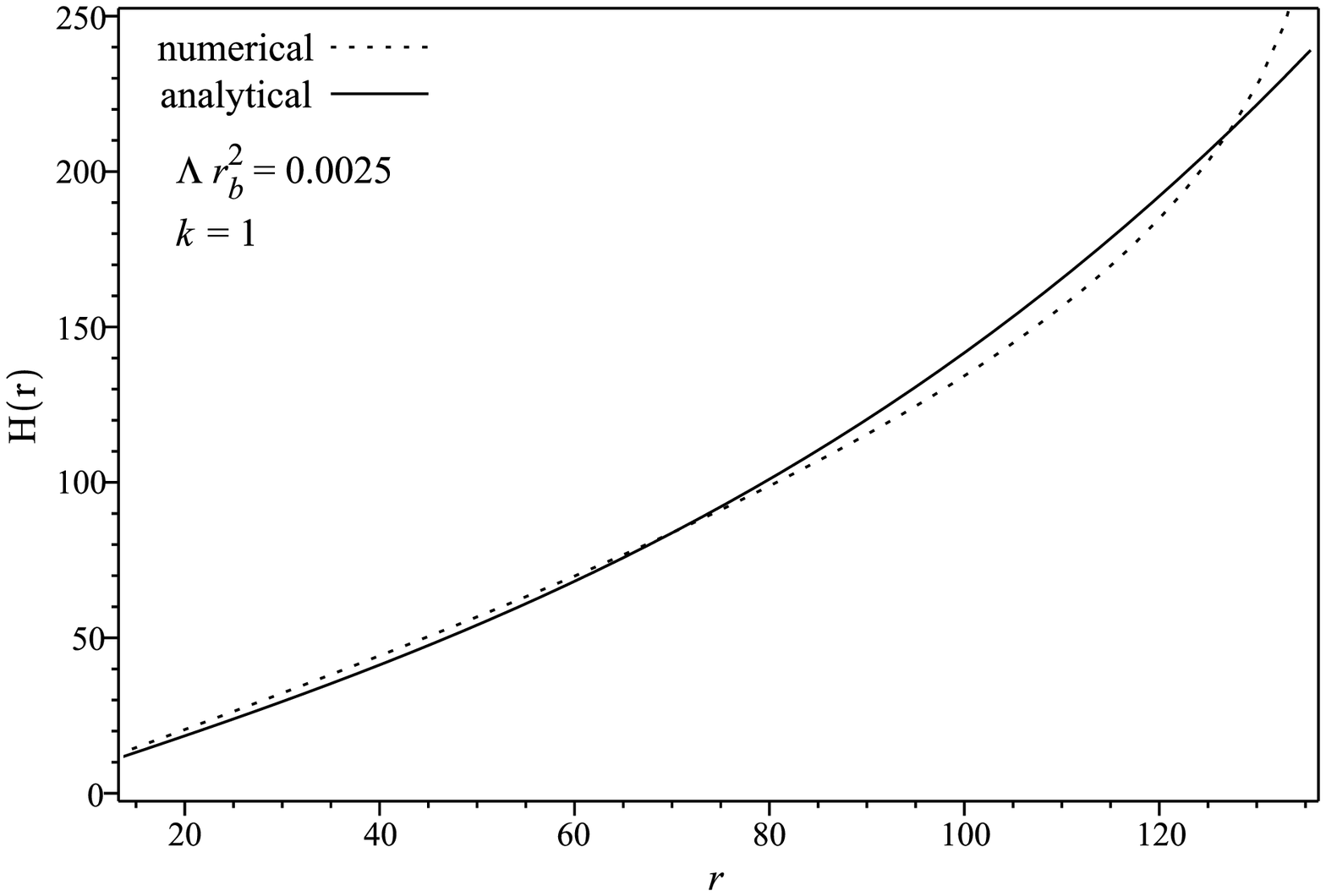}
\quad
\includegraphics[scale=0.3]{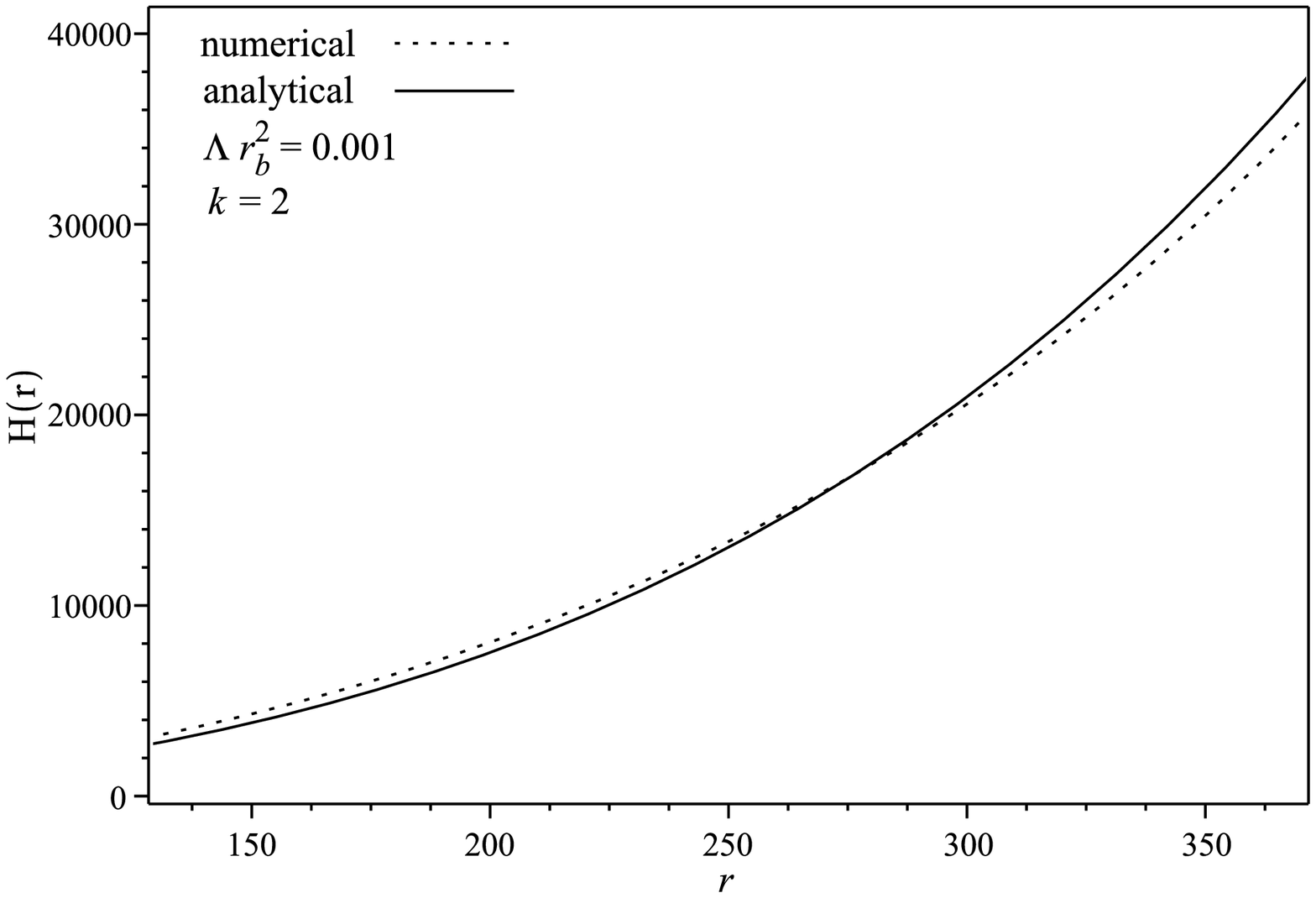}
\caption{In the left panel $r_b\approx 4$ and $r_c\approx 136.52$ while in the right panel $r_b\approx 10$ and $r_c\approx 542.65$.}
\label{f1}
\end{figure}

Inserting eq.(\ref{s14}) into eq.(\ref{s7}) we obtain a first order nonhomogeneous equation for $F^+$
\begin{equation}\label{s16}
\frac{dF^+}{dr}+\frac{k}{\sqrt{h}r}F^+=\frac{1+\lambda\sqrt{h}}{h}H_0(1+b\,I)
\end{equation}

The solution to the homogeneous part of eq.(\ref{s16}) can be obtained analytically by expanding around $h_0$ for small values of $b$, in a similar way as it was done above for $H$. This provides
\begin{equation}\label{s17}
F^+_{hom}=H_0^{-1}(1-b\,I)
\end{equation}

We find out a particular solution to the nonhomogeneous equation using the method of variable coefficient, i.e. by assuming that $F^+_{nonhom}=C(r)F^+_{hom}$, where $C(r)$ is found to be given by the following integral
\begin{equation}\label{s18}
C(r)=\int \left(\frac{1-\sqrt{h_0}}{1+\sqrt{h_0}}\right)^{-2k}\left(\frac{1+b\,I}{1-b\,I}\right)\left( \frac{1}{h}+\frac{\lambda}{\sqrt{h}} \right)\frac{r^2\,dh}{2M-2\Lambda/3\,r^3}
\end{equation}

Thus the general solution of Eq.(\ref{s6}) reads now as
\begin{equation}\label{s19}
F^+_I=\left(A_2+B_2\,C(r)\right) F^+_{hom}
\end{equation}
where $A_2$ and $B_2$ are the two integration constants.
In the next section we will match these solutions two by two in order to find an analytical expression for the greybody factors that characterize the emission of fermions by Schwarzschild-de Sitter black holes.

\section{Greybody factors and energy spectrum}

Let us start by making the observation that the matching of the solutions works only in the low-energy regime for which we can expand for small values of $\epsilon$ the near horizon solutions given in (\ref{s4}) and (\ref{s5}) to obtain
\begin{equation}\label{g1}
F^+_b\approx A^{tr} (1-i\epsilon p\ln{h}+...)
\end{equation}
\begin{equation}\label{g2}
F^+_c\approx A^{in} (1-i\epsilon q\ln{h}+...)+A^{out} (1+i\epsilon q\ln{h}+...)
\end{equation}

In order to match $F^+_b$ with $F^+_I$, respectively $F^+_I$ with $F^+_c$ we first need to find out the shape of the function $F^+_I$ in the vicinity of the two horizons. The $r\rightarrow r_i$ limit of $F^+_I$ when $r_i$ approaches the two horizons
is given by the following formula
\begin{equation}\label{g3}
\lim_{r\to r_i}F^+_I=\alpha_i\,(A_2+\beta_i\,B_2\ln h )
\end{equation}
where the constants $\alpha_i$ and $\beta_i$ are defined as
\begin{equation}\label{g4}
\begin{split}
&\alpha_i= H_0^{-1}(1-b\,I)\bigg|_{r=r_i} \\
&\beta_i=\left(\frac{1-\sqrt{h_0}}{1+\sqrt{h_0}}\right)^{-2k}\left(\frac{1+b\,I}{1-b\,I}\right)\frac{r^2}{2M-2\Lambda/3\,r^3} \bigg|_{r=r_i}
\end{split}
\end{equation}

Thus, the $r\rightarrow r_b$ limit of (\ref{s19}) can be expressed as
\begin{equation}\label{g5}
F^+_I\approx \alpha_b\,A_2+\alpha_b\beta_b\,B_2\ln h
\end{equation}

Comparing (\ref{g1}) with (\ref{g5}) will give us the following matching conditions
\begin{equation}\label{g6}
A_2=\frac{1}{\alpha_b}A^{tr} \ \ \ \ \ \ \ \ \ B_2=-\frac{i\epsilon p}{\alpha_b\beta_b}A^{tr}
\end{equation}

Taking now the $r\rightarrow r_c$ limit of (\ref{s19}) yields
\begin{equation}\label{g7}
F^+_I\approx \alpha_c\,A_2+\alpha_c\beta_c\,B_2\ln h
\end{equation}
and comparting this time with the solution at the cosmological horizon given in (\ref{g2}) will lead us in the end at
\begin{equation}\label{g8}
A^{in}=\frac{\alpha_c}{2}\left(A_2 - \frac{\beta_c}{i\epsilon q}B_2\right)\ \ \ \ \ \ \ \ \ A^{out}=\frac{\alpha_c}{2}\left(A_2 + \frac{\beta_c}{i\epsilon q}B_2\right)
\end{equation}

Taking into account the form of the asymptotic solutions (\ref{s4})-(\ref{s5}) and the conservation of flux we can now compute using relations (\ref{g6}) and (\ref{g8}) the greybody factors for which we found the following expression
\begin{equation}\label{g9}
\Gamma_j(\epsilon)\equiv 1-\bigg|\frac{A^{out}}{A^{in}}\bigg|^2=1-\left( \frac{p\,\beta_c - q\,\beta_b}{p\,\beta_c + q\,\beta_b} \right)^2
\end{equation}

The above form of the greybody factors looks very similar with the one obtained in \cite{2,15} for the case of scalar particle emission, for which the authors have also found $\Gamma_s$ to be constant at low-energies. We see that this feature remains valid also in the case of low-energy fermion emission by a SdS black hole. Moreover for fermions the greybody factor (or equivalently the absorbtion probability) has a constant value for each mode at very low-energies, for which the above (\ref{g9}) formula is valid. However, for fermions $\Gamma_j$ has a much more complicated dependence on $r_b$ and $r_c$ compared to the scalar case for which\cite{2,15}
\begin{equation}\label{g10}
\Gamma_s=1-\frac{(r_c^2-r_b^2)^2}{(r_c^2+r_b^2)^2}
\end{equation}

Let us mention that both expressions (42) and (43) for the absorption probability take the value $0$ in the limit $\Lambda\mapsto 0$ ($\Leftrightarrow$  $b\mapsto 0$). This is due to the fact that in this limit  $r_b\mapsto 2M, r_c\mapsto\infty$ in such a way that $b r_i^3= r_i-2M$. Therefore, a ratio $\frac{r_b}{r_c}\mapsto 0$ in (43) and one gets expected result. Similarly, the second term in  (42) can be expressed as a function of ratios $\frac{\beta_c}{q}, \frac{\beta_b}{p}$. First we have to calculate the limits  (cf. (29)):  $\lim_{b\rightarrow 0}{(b I_c)}= {1\over 2}k$, $\lim_{b\rightarrow 0}{(b I_b)}= 0$. These imply  $\lim_{b\rightarrow 0}\left(\frac{\beta_c}{q}\right)= 0^{-2k}\frac{1+{1\over 2}k }{1-{1\over 2}k }$, and $\lim_{b\rightarrow 0}\left(\frac{\beta_b}{p}\right)=1$ which provide the required answer $\lim_{\Lambda\rightarrow 0}{(\Gamma_j)}=0$.

In Table \ref{tab1} we present some numerical values of the greybody factors evaluated for the first three mods ($j=1/2, 3/2, 5/2$) using different values for $\Lambda r_b^2$. We observe from the table that for each mode the value of $\Gamma_j$ becomes higher as we increase the value of the cosmological constant $\Lambda$. Another conclusion that can be drawn is the fact that the contribution of the lowest mode $j=1/2$ to the emission spectra will be the dominant one.

\begin{table}[h]

{\begin{tabular}{@{}c|c c c c c@{}}
\hline
$\Lambda r_b^2$ & 0.001 & 0.0025 & 0.005 & 0.0075 & 0.01 \\
\hline
$\Gamma_{j=\frac{1}{2}}$ & 1.14 & 2.95 & 6.11 & 9.42 & 12.85 \\
\\
$\Gamma_{j=\frac{3}{2}}$ & $1.38\cdot10^{-5}$ & $9.07\cdot10^{-5}$ & $3.84\cdot10^{-4}$ & $9.03\cdot10^{-4}$ & $1.67\cdot10^{-3}$ \\
\\
$\Gamma_{j=\frac{5}{2}}$ & $1.2\cdot10^{-10}$ & $1.94\cdot10^{-9}$ & $1.56\cdot10^{-8}$ & $5.9\cdot10^{-8}$ & $1.2\cdot10^{-7}$   \\
\hline
\end{tabular}\label{tab1} }
\caption{The greybody factors for the first three modes (all the numerical values of $\Gamma_j(\epsilon)$ have been multiplied by a factor of $10^4$).}
\end{table}

Comparing numerically the greybody factors (or equivalently the absorbtion cross section) given by eq. (\ref{g9}) and (\ref{g10}) in the massless limit for the lowest angular quantum numbers ($j=1/2$ for fermions, respectively $s=0$ for scalars) we obtain that their ratio is approximatively
\begin{equation}\label{g11}
\frac{\Gamma_{j=\frac{1}{2}}}{\Gamma_{s=0}}\propto\frac{\sigma^{abs}_{j=\frac{1}{2}}}{\sigma^{abs}_{s=0}}\approx\frac{1}{12}
\end{equation}

For pure Schwarzschild black holes this ratio was showed in ref. \cite{1} to be equal to $1/8$. Thus we see that in the presence of a cosmological constant a SdS black hole will emit even more low-energy scalar quanta as Hawking radiation compared with the radiation emitted by a pure Schwarzschild black hole.

Let us now address the problem of energy spectrum. It is known that the energy emission rate for spin $1/2$ particles for a SdS black hole has the usual expression\cite{3}
\begin{equation}\label{es1}
\frac{dE}{d\epsilon}=\frac{1}{2\pi v^2}\sum_{j}(2j+1)\frac{\epsilon\,\Gamma_j}{\exp(\epsilon/T_H)+1}
\end{equation}
where $v$ represents the speed of the fermion and $T_H$ is the Bousso-Hawking temperature\cite{13} of the black hole
\begin{equation}\label{es1a}
T_H=\frac{1}{h(r_0)}\frac{1}{4\pi}\left( \frac{1}{r_b}-\Lambda r_b \right)
\end{equation}
with $r_0=(3M/\Lambda)^{1/3}$ obtained from the zero point of the first derivative of $h(r)$ .

\begin{figure}[h!t]
\includegraphics[scale=0.3]{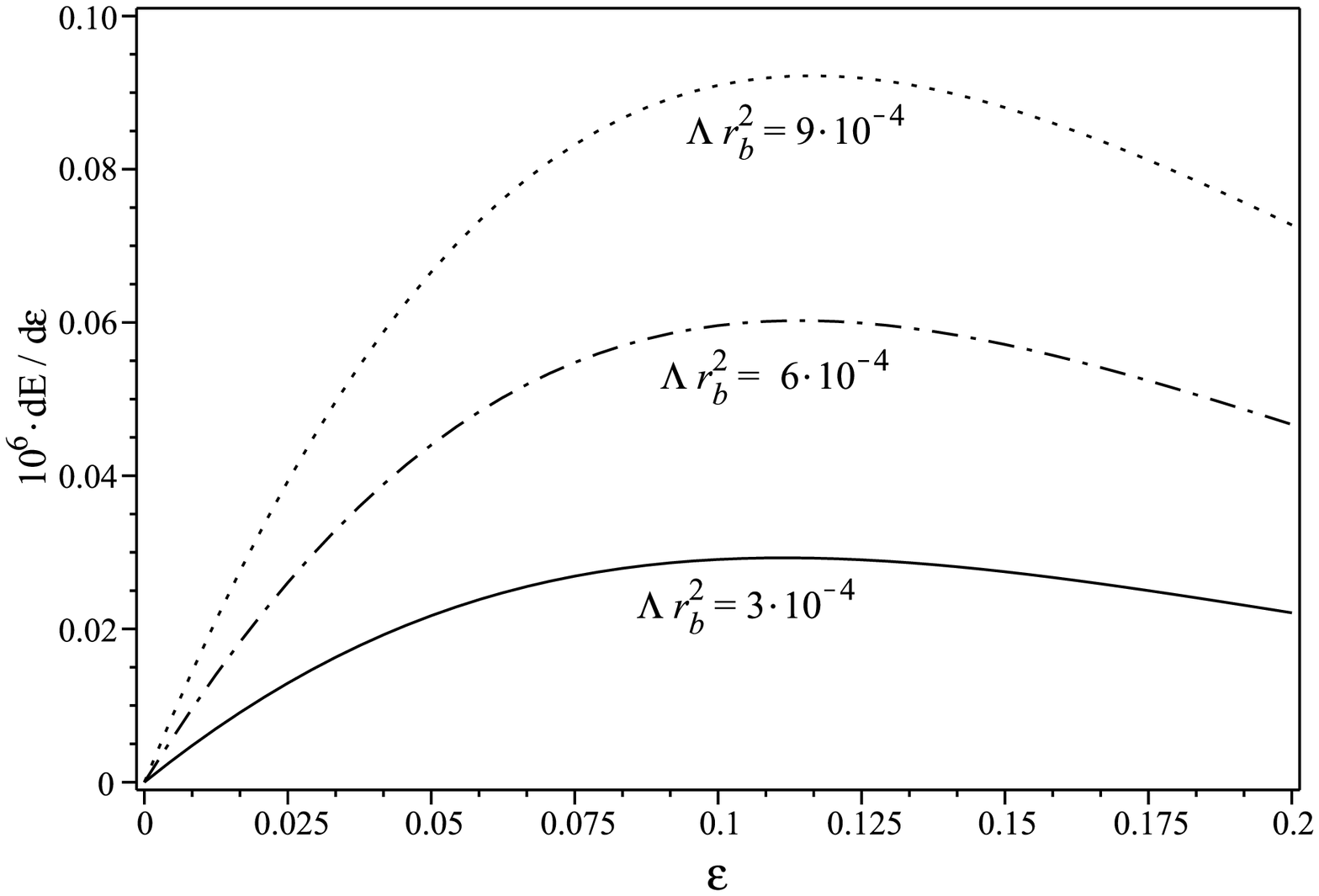}
\quad
\includegraphics[scale=0.3]{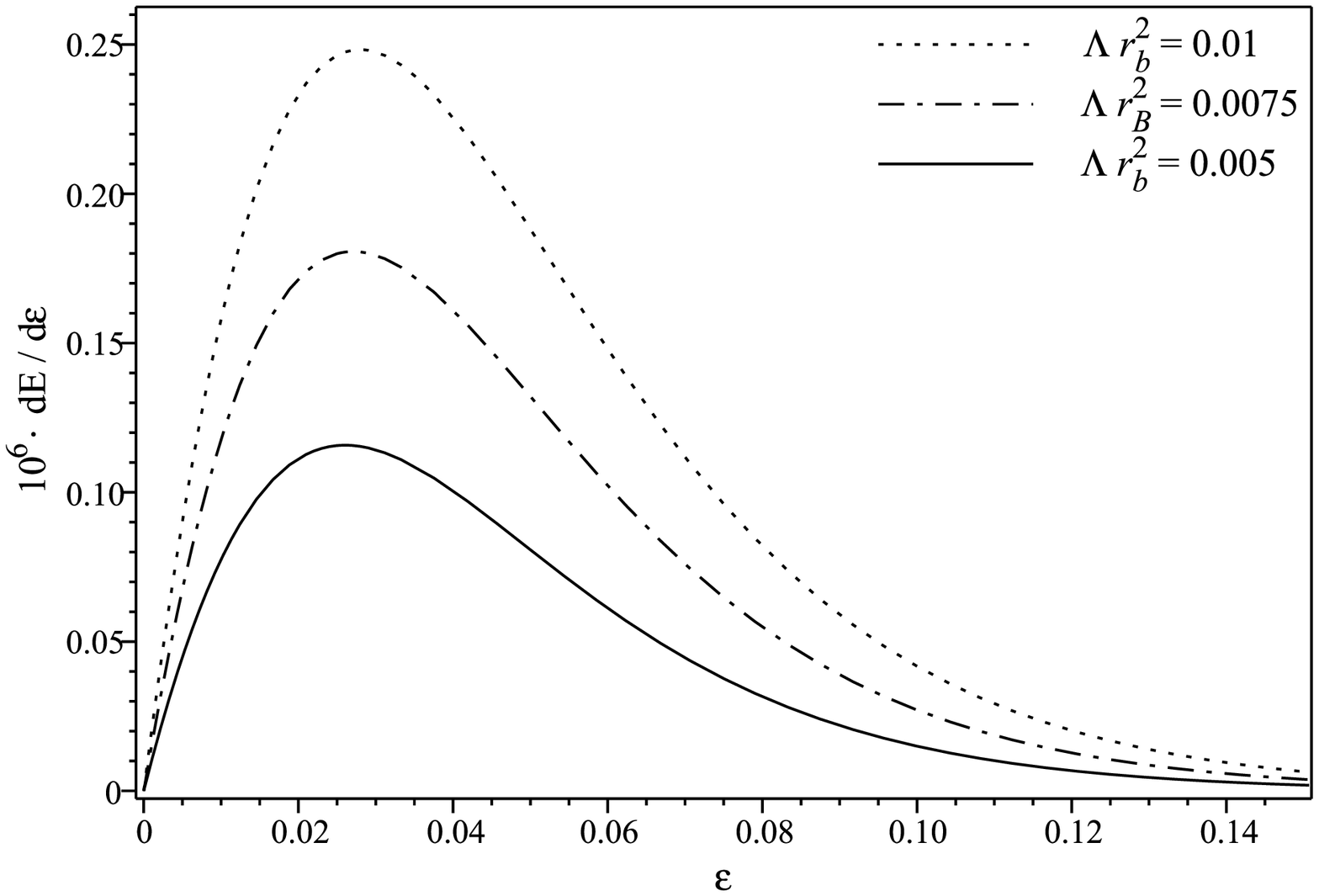}
\caption{The fermion differential energy emission rate for different values of $\Lambda r_b^2$. For the left panel we have set $r_b=1$, respectively $r_b=5$ for the right panel.}
\label{f2}
\end{figure}

The energy spectrum obtained using (\ref{es1}) and (\ref{g9}) is displayed in Fig.\ref{f2}. This spectra should be trusted for quantitative results only in the low-energy regime. In the intermediate and high-energy regimes the spectra is expected to suffer modifications due to the fact that in these regimes the greybody factor will eventually develop a dependence on energy.

We observe an enhancement of the spectrum with the increasing value of the cosmological constant. In the case of asymptotically flat Schwarzschild black holes the energy emission rate for fermions vanishes in the limit $\epsilon\rightarrow0$. We see that this remains valid also for SdS black boles. However, for massless scalar particles it turns out that this limit has a non-vanishing value for SdS as reported in ref. \cite{15,15a,15b,3}. Taking the limit of (\ref{es1}) as the energy goes to zero one can see that as long as the greybody factor has a constant value the energy emission rate will always go to zero regardless of the fermion mass. In the case of massive scalars it was showed in ref.\cite{15b} that the energy emission rate again goes to zero for a SdS black hole.

We should point out that our analytical results obtained here for low-energies are in good agrement with the results reported in \cite{3} where the authors have numerically investigated the problem for all types of fields (scalars, fermions, bosons and gravitons).

\section{Conclusions}

In this paper we calculated for the first time, analytically, the low-energy greybody factors for fermions emitted by a Schwarzschild-de Sitter black hole with a small positive cosmological constant $\Lambda$. We have found that the fermion greybody factors can be approximated by constant quantities at low energies. This result is similar to the one derived for scalar particles \cite{2,15}. Our analytical results presented here seems to confirm the numerical studies reported in Ref.\cite{3} for very small energies. Another important aspect encapsulated in our results is the fact that the ratio fermions to scalars (which is approximatively 1/12) emitted by a SdS black hole is smaller than the same ratio\cite{1} (1/8) in the case of a Schwarzschild black hole.

This work can be continued in a number of ways such as: taking into account the next order terms in solving the equation in the intermediate region presented in section 3.2 and thus possible obtaining the energy dependence of the greybody factors; finding more accurate  solutions in the regions near the black hole and cosmological horizons and thus refining the results presented here.

\section*{Acknowledgements}

C.A. Sporea was supported by the strategic grant POSDRU/159/1.5/S/137750, and in part by  a grant of the Romanian National Authority for Scientific Research, Programme for research-Space Technology and Advanced Research-STAR, project nr.  72/29.11.2013. AB acknowledges financial support from the Polish NCN project DEC-2013/09/B/ST2/03455.


\begin{thebibliography}{50}

\bibitem{20} A. G. Riess et al., Astron. J. 116, 1009 (1998).

\bibitem{20a} S. Perlmutter et al., Astrophys. J. 517, 565 (1999).

\bibitem{21} A. H. Guth, Phys. Rev. D 23, 347 (1981).

\bibitem{21a} A. D. Linde, Phys. Lett. B, 108 (6), 389-393 (1982).

\bibitem{CF} S. Capozziello, M. Francaviglia, Gen.Rel.Grav. 40 (2008) 357-420.

\bibitem{ABF} G. Allemandi, A. Borowiec, M. Francaviglia, Phys.Rev. D70 (2004) 103503.

\bibitem{FFV} M. Ferraris, M. Francaviglia, I. Volovich, Class.Quant.Grav. 11 (1994) 1505-1517.

\bibitem{AFRT} G. Allemandi, M. Francaviglia  , M.L. Ruggiero, A. Tartaglia  Gen.Rel.Grav. 37 (2005) 1891-1904.

\bibitem{22} S. W. Hawking, Commun. Math. Phys. 43, 199 (1974).

\bibitem{1} W.G. Unruh, Phys. Rev. D 14 (1976) 3251.

\bibitem{Kim} E. Jung, S. H. Kim, and D. K. Park, JHEP 0409 (2004) 005.

\bibitem{RS} M. Rogatko, A. Szyplowska, Phys.Rev.D79:104005 (2009).

\bibitem{2} P. R. Brady, C.M. Chambers, W. Krivan, and P. Laguna, Phys. Rev. D 55, 7538 (1997).

\bibitem{3} S.-F. Wu, S. Yin, G.-H. Yang, and P.-M. Zhang, Phys.Rev. D 78, 084010 (2008).

\bibitem{14} T. Harmark, J. Nata´rio, and R. Schiappa, Adv. Theor.Math. Phys. 14, 727 (2010).

\bibitem{15} P. Kanti, J. Grain, and A. Barrau, Phys. Rev. D 71, 104002 (2005).

\bibitem{15a} P. Kanti, T. Pappas, and N. Pappas, Phys. Rev. D 90, 124077 (2014).

\bibitem{15b} L.C.B. Crispino, A. Higuchi, E.S. Oliveira and J.V. Rocha, Phys. Rev. D 87, 104034 (2013)

\bibitem{16} R. Bousso and S. W. Hawking, Phys. Rev. D 57, 2436 (1998)

\bibitem{23} M. Cvetič and F. Larsen, Phys. Rev. D 57, 6297.

\bibitem{17} R. L. Mallett, Phys. Rev. D 33, 2201 (1986);

\bibitem{17a} P. C.W.Davies, L. H. Ford, and D. N. Page, Phys. Rev. D 34, 1700 (1986);

\bibitem{17b} W. H. Huang, Classical Quantum Gravity 9, 1199 (1992).

\bibitem{18} A. J. M. Medved, Phys. Rev. D 66, 124009 (2002).

\bibitem{19} C. Molina, Phys. Rev. D 68, 064007 (2003).

\bibitem{19a} R. A. Konoplya, Phys. Rev. D 68, 124017 (2003).

\bibitem{19b} V. Cardoso, O. J. C. Dias and J. P. S. Lemos, Phys. Rev. D 70, 024002 (2004).

\bibitem{6} I. I. Cotaescu, Mod. Phys. Lett. A 13, 2923 (1998).

\bibitem{6a} I. I. Cotaescu, Phys. Rev. D 60, 124006 (1999).

\bibitem{7} B. Thaller, The Dirac Equation (Springer Verlag, Berlin Heidelberg, 1992).

\bibitem{8} V. B. Berestetski, E. M. Lifshitz and L. P. Pitaevski, Quantum Electrodynamics (Pergamon Press, Oxford 1982).

\bibitem{9} H. Y. Liu, Gen. Rel. Grav.23, 759 (1991).

\bibitem{10} Z. Stuchlik and S. Hledik, Phys. Rev. D 60, 044006 (1999).

\bibitem{11} H. Nariai, Sci. Rep. Tohoku Univ. 34, 160 (1950).


\bibitem{4} R. M. Wald, Commun. Math. Phys. 45, 9 (1975).

\bibitem{5} S.W. Hawking, Phys. Rev. D 14, 2460 (1976).

\bibitem{12} G.W. Gibbons and S.W. Hawking, Phys. Rev. D 15, 2738(1977).

\bibitem{13} R. Bousso and S.W. Hawking, Phys. Rev. D 54, 6312 (1996).



\end{thebibliography}
\end{document}